\documentclass[letterpaper]{article} % DO NOT CHANGE THIS
\usepackage[]{aaai25}  % DO NOT CHANGE THIS
\usepackage{aaai25}
\usepackage{times}  % DO NOT CHANGE THIS
\usepackage{helvet}  % DO NOT CHANGE THIS
\usepackage{courier}  % DO NOT CHANGE THIS
\usepackage[hyphens]{url}  % DO NOT CHANGE THIS
\usepackage{graphicx} % DO NOT CHANGE THIS
\usepackage{amsmath}
\usepackage{amssymb}
\usepackage{subcaption}
\usepackage{booktabs}
\usepackage{subfloat}

\usepackage{natbib}  % DO NOT CHANGE THIS AND DO NOT ADD ANY OPTIONS TO IT
\usepackage{caption} % DO NOT CHANGE THIS AND DO NOT ADD ANY OPTIONS TO IT
\usepackage{amsmath}
\usepackage{amssymb}
\usepackage{algpseudocode}
\usepackage{algorithmicx,algorithm}
\usepackage{color}
\definecolor{S1}{RGB}{45,86,174}
\definecolor{S11}{RGB}{42,80,162}
\definecolor{S12}{RGB}{29,45,112}
\definecolor{S13}{RGB}{20,71,83}
\definecolor{S2}{RGB}{41,22,140}
\definecolor{S3}{RGB}{0,124,130}

\usepackage{newfloat}
\usepackage{listings}
\DeclareCaptionStyle{ruled}{labelfont=normalfont,labelsep=colon,strut=off} % DO NOT CHANGE THIS
\floatstyle{ruled}
\newfloat{listing}{tb}{lst}{}
\floatname{listing}{Listing}
\makeatletter

\newcommand{\Rmnum}[1]{\expandafter\@slowromancap\romannumeral #1@}
\makeatother

\title{ERFSL: An Efficient Reward Function Searcher via Language Models for Custom-Environment Multi-Objective Optimization  (Student Abstract)}
\author{
    Guanwen Xie\textsuperscript{\rm 1},  
    Jingzehua Xu\textsuperscript{\rm 1},
    Yiyuan Yang\textsuperscript{\rm 2},
    Yimian Ding\textsuperscript{\rm 1}\equalcontrib,
    Shuai Zhang\textsuperscript{\rm 3}\equalcontrib
}
\affiliations{
    \textsuperscript{\rm 1}Tsinghua Shenzhen International Graduate School, Tsinghua University, China\\
    \textsuperscript{\rm 2}Department of Computer Science, University of Oxford, United Kingdom\\
    \textsuperscript{\rm 3}Department of Data Science, New Jersey Institute of Technology, USA\\
    gwxie360@outlook.com,
    xjzh23@mails.tsinghua.edu.cn,
    yiyuan.yang@cs.ox.ac.uk,
    yimiandingthu@gmail.com,
    sz457@njit.edu
}

\begin{document}

\maketitle

\begin{figure*}[!t]
    \centering
    \includegraphics[width=0.803\linewidth]{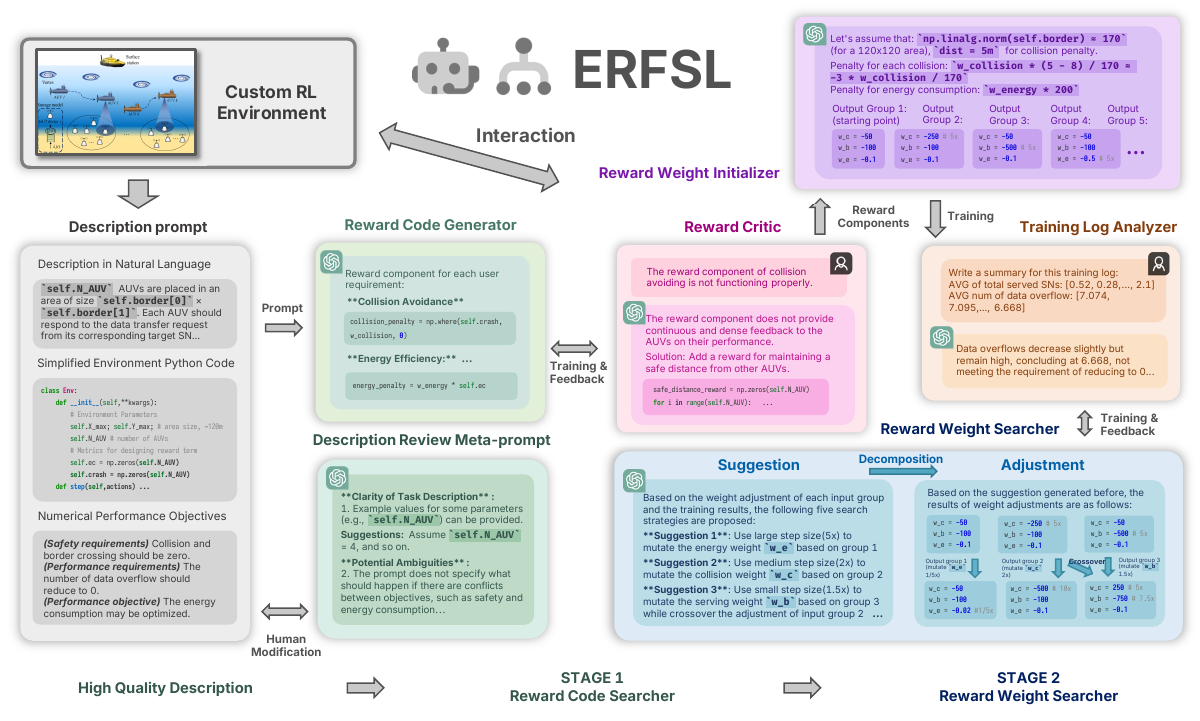}
    \caption{The main architecture and prompt examples of the proposed ERFSL.}
    \label{fig_1}
\end{figure*}

% Uncomment the following to link to your code, datasets, an extended version or similar.
%
% \begin{links}
%     \link{Code}{https://aaai.org/example/code}
%     \link{Datasets}{https://aaai.org/example/datasets}
%     \link{Extended version}{https://aaai.org/example/extended-version}
% \end{links}
\section{Abstract}\label{se:0}
We propose ERFSL, an efficient reward function searcher using large language models (LLMs) for custom-environment, multi-objective learning-based methods (LB). ERFSL generates reward components based on explicit user requirements, rectifies them using a reward critic, and iteratively optimizes the weights of these components based on textual context generated by the training log analyzer. Applied to a simulation-based benchmark task, the reward critic corrects reward codes with only one feedback iteration per requirement, and the reward weight initializer acquires diverse reward functions within the Pareto set. Even when a weight is off by a factor of 500, an average of only 5.2 iterations is needed to meet user requirements. The approach works adequately with GPT-4o mini and does not require advanced understanding capabilities. The full-text prompts, examples of LLM-generated answers, and study files are available at https://360zmem.github.io/LLMRsearcher/ .
    
\section{Introduction}\label{se:1}
Learning-based modeling(LB) is useful for multi-objective tasks; however, designing complex reward functions remains challenging due to ambiguous and varied requirements\cite{4}. Large language model (LLM)-aided reward function design has demonstrated remarkable performance in various scenarios, such as customized control tasks \cite{2,8,17} and virtual game play \cite{5,7}, yet issues such as incorrect code and imbalanced weights may arise with intricate tasks. To address these challenges, some approaches decompose complex tasks into several sub-tasks or skills, while providing clear task feedback accordingly \cite{21,22,23}. 

In this paper, we decompose the LLM-aided reward function generation into reward component design and weight assignment, employing LLMs as white-box searchers under textual context and clear feedback to fully leverage their semantic understanding capabilities. The ERFSL's architecture and key prompts are shown in Figure 1.

\section{Methodology}\label{se:3}

\textbf{Environment Description.} Task description is a common part of most subsequent prompts, including text descriptions, environment code or APIs, and user requirements. We decompose user requirements into specific numerical objectives (e.g., achieving zero violations in constraint handling). To avoid ambiguous descriptions and facilitate human modification, we design a meta-prompt to allow LLMs to enhance the description quality.

\textbf{Reward Code Generation.} We borrow the LLM-aided reward code design frameworks, creating a reward component for each user requirement. However, the initial LLM-generated code may be incorrect due to the absence of prior knowledge about the environment and complex contexts. Therefore, we test each component separately and correct errors using the reward critic. LLMs can also fabricate necessary variables and prompt users to complete them, handling incomplete environment descriptions effectively.

 % 实验环境概述。我们利用
\textbf{Reward Weight Search.} Multi-objective learning-based methods require a balanced scale of reward components. We first utilize a reward weight initializer to designate K=5 groups of weights to ensure the components' values are on the same scale by pre-calculating the values of reward components. After that, LLMs suggest K=5 weight adjustments and generate weight groups based on the training results summarized by the training log analyzer. To prevent ambiguity and potential redundancies in weight adjustments across multiple input groups, inspired by genetic algorithms, the searcher specifies the starting point and adjustment direction (increase/decrease/fine-tune) to mutate (adjust) each weight. For multiple weight adjustments, we conduct crossovers between the mutated input groups. Additionally, we separate the process of generating adjustment suggestions and output weight groups to shorten the prompt and ensure precise understanding and execution.

\section{Experiments and Main Results}\label{se:4}
We utilize ERFSL to design reward functions in a zero-shot manner for a previous simulation-based benchmark task via mobile units \cite{14,26} using a standard model setting, considering that domain tasks may present more challenges and possess more user requirements \cite{25}. We primarily use gpt-4o-2024-08-06 (denoted as \textbf{GPT-4o}), and also conduct tests with gpt-4o-mini-2024-07-18 (denoted as \textbf{GPT-4om}). Similar to Eureka \cite{2}, we design a baseline \textbf{EUREKA-M} that takes multiple full reward functions along with their training logs as input and outputs K=5 reward functions.

\begin{table}[t]
            \centering
            \caption{The number of iterations of searching weights under different experiment settings, and each is performed 5 times.}
            \label{table:masking_performance}
            %\vspace{3mm}
            \begin{tabular}{|l|c|c|c|} 
            \hline
            Setting & \textbf{GPT-4o} & \textbf{GPT-4o w/o TLA} & \textbf{GPT-4om}  \\
            \hline
            \textbf{RWI}    & \multicolumn{3}{|c|}{\textbf{0.40±0.49}}  \\
            \hline
            \textbf{RWI-UB}  & 1.20±0.75 & 1.60±1.02 & 1.80±1.17 \\ %TODO
            \hline
            \textbf{500x off}  & 5.20±1.46 & 6.40±1.86 & 8.60±1.74  \\ %TODO
            \hline
            \end{tabular}
            %\vspace{-1mm}
            \end{table}

\begin{figure}[!t]
    \centering
    \includegraphics[width=0.495\linewidth]{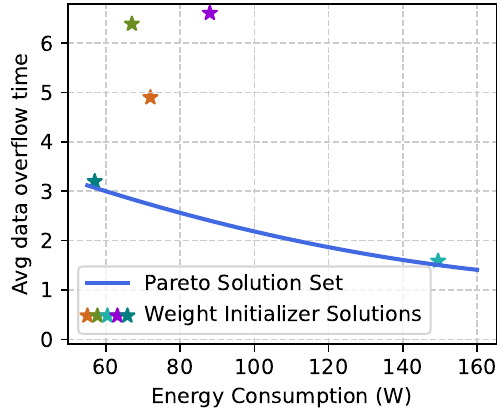}    \includegraphics[width=0.495\linewidth]{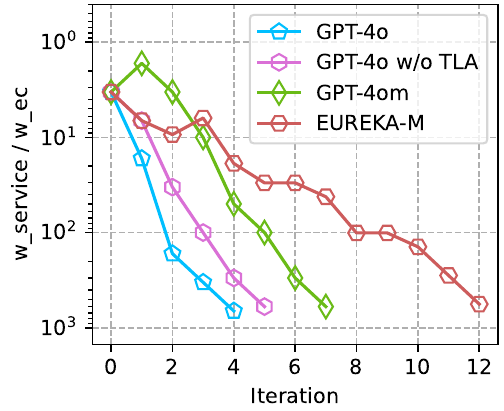}
    \caption{(a) Solutions generated from the reward weight initializer. (b) Change of maximum value of w\_service/w\_ec under different settings during iteration.}
    \label{fig_2}
\end{figure}
% reward critic可以在仅需一次反馈下即可纠正代码错误，而这些是EUREKA-M等基线难以做到的，避免了难以纠正的错误，这得益于任务拆分所带来的明确反馈。
\vspace{4mm}
\textbf{The reward critic can correct code errors with just one feedback per component,} which avoids errors that are hard to correct when utilizing baselines like EUREKE-M, thanks to the explicit feedback coming with task splitting.

% 表1展示了各种实验设定下达成用户需求所需的搜索次数，图2(a)展示了reward weight initializer生成的初始权重组(denoted as \textbf{RWI})，Fig. 2(b) shows the maximum value of w_service/w_ec in five weight groups during training, with the best performance among the five repetitions。可以看出经过reward weight initializer的初始化，五组解中有两组实现了Pareto解，without search or only with a refined search。即使初始化权重过程中去除了数值参考和平衡要求(denoted as \textbf{RWI-UB}),也只需要0-3次搜索即可达成用户需求。当能耗权重被增大500倍时(denoted as \textbf{500x off})，reward weight searcher成功发现问题并采取了灵活的步长策略，仍然仅需要平均5.2次调整即可达到用户需求。去除掉training log analyzer(denoted as \textbf{GPT-4o w/o TLA})之后，灵活性有所降低。另外EUREKA-M调节策略展现了相当的随机性，且以随机尝试提升权重为主，缺乏针对性。
\textbf{Reward weight initialization and search.} Table 1 presents the number of searches required to meet user requirements. Figure 2(a) illustrates the initial weight groups generated by the reward weight initializer (denoted as \textbf{RWI}), while Figure 2(b) depicts the maximum value of w\_service/w\_ec in five weight groups during training, demonstrating the best performance among the five repetitions. Two of the five sets of solutions generated by the RWI achieve Pareto solutions. Even if the numerical reference and balance requirements (denoted as \textbf{RWI-UB}) are removed from the prompt of RWI, only 0-3 searches are required to meet user requirements. When the energy weight is increased by a factor of 500 (denoted as \textbf{500x off}), the reward weight searcher successfully identifies the problem and adopts a flexible step size strategy, requiring only an average of 5.2 adjustments. Flexibility is reduced when the training log analyzer (denoted as \textbf{GPT-4o w/o TLA}) is removed. Moreover, EUREKA-M adopts small, random step sizes and increases weights in a highly random manner rather than decreasing the penalty for energy consumption, necessitating a substantial number of iterations.

% 由于较短上下文和the process of reward weight searcher的拆分，从而避免了small-scale LLMs的劣势场景，(尽管内容生成prompt的回答质量略有下降，但)除了RWI之外的所有过程，包括the reward code generation和reward weight search均表现adequate。尽管步长的选取不灵活，但性能仍然超越了EUREKA-M。
\vspace{5.5mm}
\textbf{The performance of GPT-4om.} The splitting of the reward weight search process shortens the context, thereby avoiding the disadvantages faced by small-scale LLMs \cite{24}. Although the answer quality of the content generation prompt is slightly degraded and the reward weight initializer not functions, the performance of reward code generation and reward weight search is adequate. Although the step size selection lacks flexibility, the final performance still surpasses that of EUREKA-M.

% In conclusion，受益于用户需求的拆分和明确的任务反馈，ERFSL能够生成正确的reward components；而文本上下文反馈和多输入输出搜索策略使得较少的权重调节次数即可平衡这些reward components。对于reward weight search搜索流程的拆分降低了对于模型理解能力的需求，增大了适用性。(未来工作照抄ICASSP)。
% 不要结论了，直接提前。future work一并因此受到影响，也不打算要了
% In conclusion, ERFSL can generate correct reward components thanks to the clear split of user requirement and feedback. Textual context and multiple input/output search strategies balance these reward components with fewer weight adjustments. Future work will focus on automating task descriptions and verifying the verifying the framework across various tasks. 
\vspace{6mm}
\section{Acknowledgment}

Part of this work was done when Guanwen Xie and Jingzehua Xu were studying in the MicroMasters Program in Statistics and Data Science at Massachusetts Institute of Technology (MIT). We are very grateful to Yiyuan Yang and Shuai Zhang at University of Oxford and New Jersey Institute of Technology (NJIT) for their strong support, and to Miao Liu and Songtao Lu at IBM research and MIT-IBM Watson AI Lab and for their valuable advice, respectively. Additionally, we thank all anonymous reviewers for their constructive comments.

\section{Impact Statement}

This paper presents work whose goal is to explore using LLMs to design reward functions for the multi-objective tasks. There are many potential societal consequences of our work, none of which we feel must be specifically highlighted here.

\bibliography{aaai25}

      \end{document}